\documentclass[prc,onecolumn,showpacs,superscriptaddress]{revtex4}
\usepackage{graphicx,amssymb,amsmath}

\begin{document}

\title{The $\beta^4$ potential at the U(5)--O(6) critical point of the Interacting Boson Model}

\author{Jos\'e Enrique Garc\'{\i}a-Ramos}
\email{enrique.ramos@dfaie.uhu.es}
\affiliation{Departamento de F\'{\i}sica Aplicada, Universidad de
Huelva, 21071 Huelva, Spain}

\author{Jorge Dukelsky}
\email{dukelsky@iem.cfmac.csic.es}
\affiliation{Instituto de Estructura de la Materia, CSIC, Serrano
123, 28006 Madrid, Spain}

\author{Jos\'e M. Arias}
\email{ariasc@us.es}
\affiliation{Departamento de F\'{\i}sica At\'omica, Molecular y
Nuclear, Facultad de F\'{\i}sica, Universidad de Sevilla,
Apartado~1065, 41080 Sevilla, Spain}

\bigskip

\begin{abstract}
Exact numerical results of the interacting boson model Hamiltonian
along the integrable line from U(5) to O(6) are obtained by
diagonalization within boson seniority subspaces. The matrix
Hamiltonian reduces to a block tridiagonal form which can be
diagonalized for large number of bosons. We present results for
the low energy spectrum and the transition probabilities for
systems up to 10000 bosons, which confirm that at the critical
point the system is equally well described by the Bohr Hamiltonian
with a $\beta^4$ potential.
\end{abstract}


\maketitle

The main goal of this note is to report on new results that
complete a previous study \cite{beta4} on the relations between
the critical point in the transition from U(5) to O(6) limits of
the interacting boson model (IBM) \cite{IBM} and the recently
proposed E(5) critical point symmetry \cite{E5}. In Ref.
\cite{beta4} two different boson Hamiltonians performing the
transition from U(5) to O(6) were used to show that at the
critical point: i) they provide different spectra and transitions
for small number of bosons, ii) they converge to the same spectrum
for large N and iii) both converge to the spectrum provided by the
Bohr Hamiltonian \cite{BMII} with a $\beta^4$ potential rather
than to the one provided by a square well potential as in the E(5)
model. Most of the large N analysis was based on the solution of
the Richardson equations \cite{Duk01,Richard} which allow to
obtain energy eigenvalues but the form of the eigenstates is not
well suited to calculate transition probabilities. Therefore, in
order to study transition rates we had to resort to current IBM
codes that restricted our calculations to systems up to N=40
\cite{beta4}. For these small N values, the transition rates show
a tendency to approach the $\beta^4$ potential results but they are not
conclusive. In this brief report we present an alternative to the
Richardson equations for obtaining the low energy eigenvalues and,
on the same footing, the transition probabilities in the
U(5)--O(6) transitional region for large N values.

The two Hamiltonians describing the U(5)--O(6) transition studied
in Ref.\cite{beta4} are

\begin{equation}
\hat H_I = x \hat n_d + \frac{1-x}{N-1} \hat P^\dag \hat P ~,
\label{HPP}
\end{equation}
and
\begin{equation}
\hat H_{II} = x \hat n_d - \frac{1-x}{N} \hat Q^{\chi=0} \cdot \hat
Q^{\chi=0} ~,
\label{HQQ}
\end{equation}
where
\begin{eqnarray}
\label{nd}
\hat n_d &=& \sum_m d^\dag_m d_m , \\
\label{P}
\hat P^\dag&=&\frac{1}{2} ~ (d^\dag \cdot d^\dag -
s^\dag \cdot s^\dag)=\frac{1}{2} \left(P^\dag_d-P^\dag_s\right), \\
\label{Q}
\hat Q^{\chi=0}&=& (s^{\dagger}\times\tilde d
+d^\dagger\times\tilde s)^{(2)},
\end{eqnarray}
and $\cdot$ stands for the scalar product. We have introduced in
(\ref{P}) the boson pair creation
operators $P^\dag_d=d^\dag \cdot d^\dag$ and
$P^\dag_s=s^\dag \cdot s^\dag$ that will be used later on.

The mean field analysis of the quantum phase diagram of the IBM is
usually performed within the intrinsic state formalism
\cite{Gino80,Diep80} where, after separating the three Euler
angles, the trial wave function is a boson condensate depending on
the two geometrical variables $\beta$ and $\gamma$. Along the
U(5)--O(6) transition the energy surface is $\gamma-$independent and
the intrinsic ground state energy for a given value of the control
parameter $x$ corresponds to the value of the deformation parameter,
$\beta$, which minimizes the energy surface.
The phase transition along this line is then signaled by
the condition
\begin{equation}
[d^2 E (N, \beta) /d \beta^2]_{\beta=0}=0,
\end{equation}
which fixes the critical value of the control parameter $x$. For
the Hamiltonian (\ref{HPP}) the critical $x$ is $x_c^I=0.5$, independent of the
number of bosons $N$, while for the Hamiltonian (\ref{HQQ}) it is
$x_c^{II}=\frac{4N-8}{5N-8}$. In the large N limit
$x_c^{II}\rightarrow 4/5$.

In order to study the eigenstates of the Hamiltonians (\ref{HPP})
and (\ref{HQQ}) we introduce the $s-$ and $d-$boson pair algebra \cite{PanDra}
\begin{eqnarray}
K^{+}_s &=& \frac{1}{2} s^\dag \cdot s^\dag=\frac{1}{2} P^\dag_s=
\left(K^{-}_s  \right)^\dagger, \nonumber \\
K_{s}^{0} &=&\frac{1}{2} \left( s^\dag s
+\frac{1}{2}\right)  =\frac{1}{2}\hat n_{s}+\frac{1}{4},
 \label{Ks} \\
K^{+}_d &=& \frac{1}{2} d^\dag \cdot d^\dag=\frac{1}{2} P^\dag_d=
\left(K^{-}_d  \right)^\dagger, \nonumber \\
K_{d}^{0} &=&\frac{1}{2} \sum_m \left( d^\dag_m  d_m
+\frac{1}{2}\right)  =\frac{1}{2}\hat n_{d}+\frac{5}{4}.
 \label{Kd}
\end{eqnarray}
For each $\ell$ value, $0$ or $2$, the three operators
$\{K^{+}_\ell,K^{-}_\ell,K_{\ell}^{0}\}$ satisfy the $su(1,1)$
commutator algebra%
\begin{equation}
\left[  K_{\ell}^{0},K_{\ell^{\prime}}^{\pm}\right]
=\pm\delta_{\ell\ell^{\prime}}%
K_{\ell}^{\pm}~,~\left[  K_{\ell}^{+},K_{\ell^{\prime}}^{-}\right]  =-2\delta
_{\ell\ell^{\prime}}K_{\ell}^{0}. \label{com2}%
\end{equation}

A complete set of eigenstates for a general IBM U(5)--O(6) transitional
Hamiltonian
can be written in terms of the raising
operator $K^{+}_\ell$ acting on a subspace of unpaired bosons
characterized by the seniority quantum number $\nu$
\begin{equation}
\left\vert \tilde n_{\ell} \nu_{\ell} \right\rangle
=\frac{1}{\sqrt{C_{\ell,\nu_\ell}^{\tilde
n_{\ell}}}}\left(K_{\ell}^+\right)^{\tilde n_{\ell}}\left\vert
\nu_{\ell}\right\rangle ,
\end{equation}
where $\nu_{s}=0,1$, $\nu_{d}=0,1,2,\dots$, and
$\vert\nu_{\ell}\rangle$ is a normalized state. The value of
$\nu_{\ell}$ gives the number of bosons of type $\ell$ not coupled
in pairs to zero angular momentum. The label $\tilde n_\ell$
refers to boson pairs coupled to zero angular momentum. Therefore,
the total number of bosons is $N=2\tilde n_s+2 \tilde n_{d}+\nu_s
+ \nu_d$. Using the $su(1,1)$ algebra it is straightforward to
obtain the normalization constants
\[
C_{\ell,\nu_\ell}^{\tilde n_{\ell}}=
\langle \nu_\ell\vert \left(K_{\ell}^-\right)^{\tilde n_{\ell}}
\left(K_{\ell}^+\right)^{\tilde n_{\ell}}
\vert\nu_{\ell}\rangle
=\frac{\tilde n_{\ell}!\left(
2\tilde n_{\ell}+2\ell+2\nu_{\ell}-1\right)  !!}{2^{\tilde
n_{\ell}}\left(  2\ell+2\nu_{\ell}-1\right)!!} .%
\]
Now we proceed to construct the complete set of states as
\begin{equation}
\left\vert \tilde n_{s}\tilde n_{d},\nu_{s}\nu_{d}\right\rangle
=\frac{1}{\sqrt{C_{s,\nu_s}%
^{\tilde n_{s}}C_{d,\nu_d}^{\tilde n_{d}}}}\left(
K_{s}^{+}\right)  ^{\tilde n_{s}}\left(  K_{d}%
^{+}\right)  ^{\tilde n_{d}}\left\vert \nu_{s}\nu_{d}\right\rangle .
\label{base}
\end{equation}
The basis (\ref{base}), although it has not good
angular momentum, is especially useful for diagonalizing
the Hamiltonians (\ref{HPP}) and (\ref{HQQ}). To show this, we rewrite
the Hamiltonians (\ref{HPP}) and (\ref{HQQ})
in terms of the generators of the two $su(1,1)$ algebras
\begin{widetext}
\begin{eqnarray}
\hat H_I &=& x \hat n_d + \frac{1-x}{(N-1)} \left(K_s^+K_s^-+K_d^+K_d^-
-K_s^+K_d^- -K_d^+K_s^- \right)~,
\\
\hat H_{II} &=& x \hat n_d - \frac{1-x}{N} \left(4K_s^+K_d^- +
  4K_d^+K_s^- + 5 \hat n_s +\hat n_d + 2 \hat n_s \hat n_d\right)~.
\end{eqnarray}
The matrix elements of the relevant operators for both Hamiltonians
in the basis (\ref{base}) are:
\begin{eqnarray}
\left\langle \tilde n_{s}\tilde n_{d},\nu_{s}\nu_{d}
\right\vert \hat n_{s}\left\vert \tilde n_{s}%
\tilde n_{d},\nu_{s}\nu_{d}\right\rangle &=& 2\tilde n_{s}+\nu_{s}
\nonumber \\
\left\langle \tilde n_{s}\tilde n_{d},\nu_{s}\nu_{d}\right\vert
\hat n_{d}\left\vert \tilde n_{s}%
\tilde n_{d},\nu_{s}\nu_{d}\right\rangle &=& 2\tilde n_{d}+\nu_{d}%
\nonumber \\
\langle \tilde n_{s}\tilde n_{d},\nu_{s}\nu_{d} \vert K_{s}^{+}K_{s}^{-}\vert
\tilde n_{s}\tilde n_{d},\nu_{s}\nu_{d}\rangle
&=& \tilde n_{s} \left(  \tilde n_{s}+\nu_{s}-\frac{1}{2}\right)
\nonumber \\
\left\langle \tilde n_{s}\tilde n_{d},\nu_{s}\nu_{d}
\right\vert K_{d}^{+}K_{d}^{-}\left\vert
\tilde n_{s}\tilde n_{d},\nu_{s}\nu_{d}\right\rangle &=&
 \tilde n_{d}\left(  \tilde n_{d}+\nu_{d}+\frac
{3}{2}\right)
\nonumber \\
\langle \left(  \tilde n_{s}-1\right)
\left(  \tilde n_{d}+1\right)  ,\nu_{s}\nu
_{d}\vert K_{d}^{+}K_{s}^{-}\vert \tilde n_{s}\tilde n_{d},\nu_{s}\nu
_{d}\rangle
&=&\frac{1}{2}\sqrt{\tilde n_{s}\left(  \tilde n_{d}+1\right)  \left(
2\tilde n_{s}+2\nu_{s}-1\right)  \left(  2\tilde
n_{d}+2\nu_{d}+5\right)  }.
\end{eqnarray}
\end{widetext}

\begin{figure}[h]
  \centering
  \includegraphics[width=12cm]{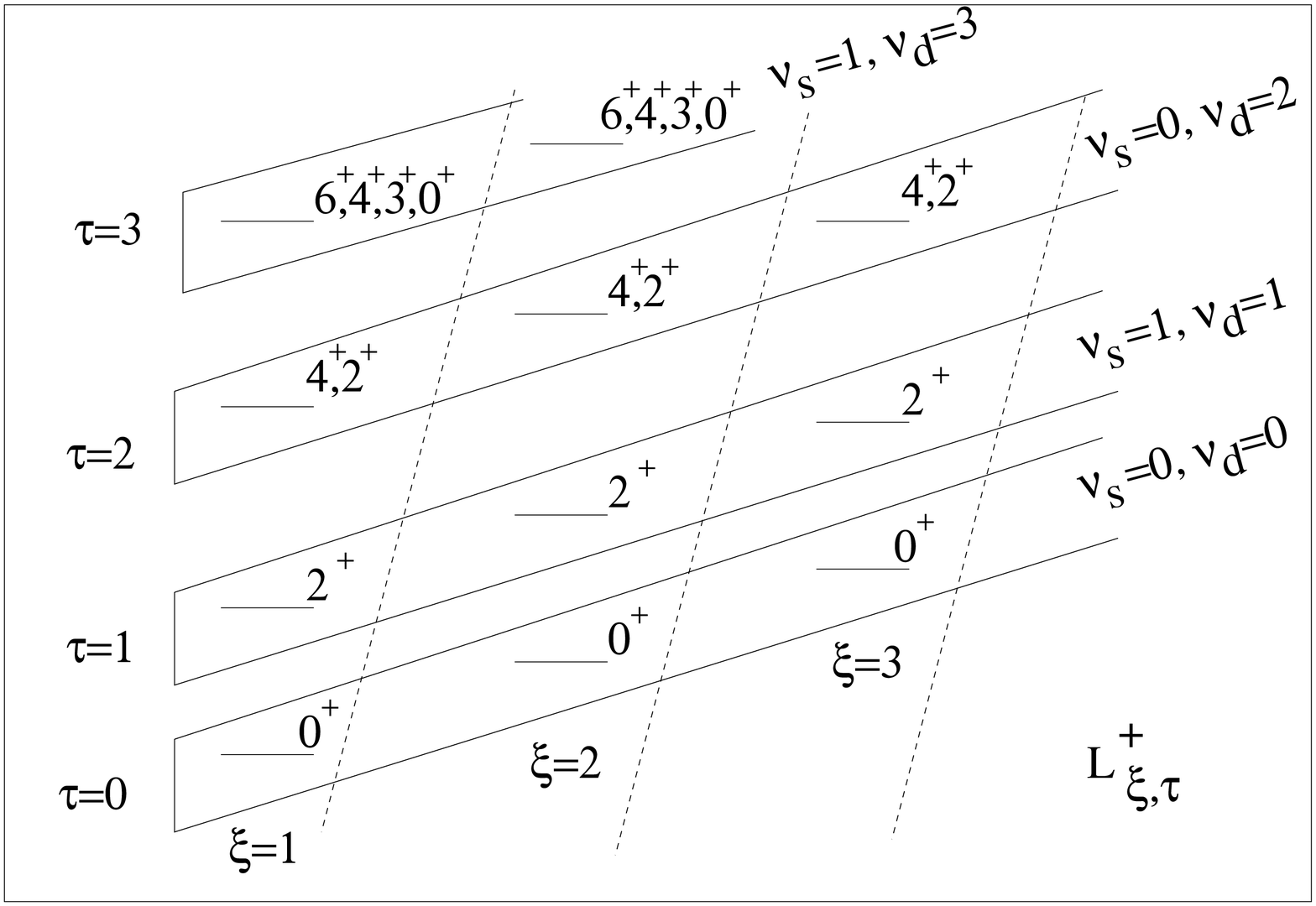}
\caption{Schematic spectrum obtained by diagonalization within boson
  seniority subspaces as explained in the text and its correspondence
  with the one of Refs.~\cite{E5} and \cite{beta4}.}
\label{fig-states}
\end{figure}

The Hamiltonians (\ref{HPP}) and (\ref{HQQ}) do not mix states
with different seniority quantum numbers ($\nu_s,\nu_d$) leaving
invariant these seniority subspaces.  Within each subspace the
Hamiltonian matrices are tridiagonal and can be easily
diagonalized for very large N values. We will label states within
each subspace by the quantum number $\xi$. It is worthwhile to
note here that $d$ boson seniority, $\nu_d$, is equivalent to the
O(5) quantum number $\tau$ \cite{IBM}. The construction of the
spectrum for a system with even number of bosons is as follows,
one starts with the subspace $\tau=0$ ($\nu_s=0,\nu_d=0$) where
all the bosons are coupled in pairs of zero angular momentum.
Consequently, states within this subspace will have total angular
momentum $L=0$.  The lowest eigenvalue ($\xi=1$) is the ground
state $0^+_{1,0}$ (the notation is $L^\pi_{\xi,\tau}$ \cite{E5}),
the second lowest eigenvalue is the first excited state $\tau=0$
$L^\pi=0^+$ state, which is labelled $0^+_{2,0}$, etc. The next
block with $\tau=1$ ($\nu_s=1,\nu_d=1$) has one pair broken into
an $s$-boson and a $d$-boson. Correspondingly, all states in this
block have $L=2$. The lowest eigenvalue ($\xi=1$)
is the lowest $2^+$ which is labelled
as $2^+_{1,1}$, the next one ($\xi=2$) is $2^+_{2,1}$, etc. The
next block is for $\tau=2$ ($\nu_s=0,\nu_d=2$) and corresponds to
one broken boson pair into two $d$-bosons. It provides states with
angular momenta $L=4,2$ (notice that $L=0$ is excluded from this
subspace since it is included in the $\nu_s=0,\nu_d=0$ subspace).
One can continue in this way with the
next block, $\tau=3$ ($\nu_s=1,\nu_d=3$). It corresponds to two boson
pairs broken into one $s-$boson and three $d-$bosons and gives rise to
$L=6,4,3,0$ states and so on. The ground state band is formed by all lowest
($\xi=1$) eigenstates for $\tau=0,1,2,3,\dots$ The first excited
band ($\xi=2$) is formed by the next lowest $\tau=0,1,2,3,\dots$
eigenstates, etc. Following this sequence one finds the well known
triangular structure associated to O(5). All this is shown
schematically in Fig.~\ref{fig-states}.

The diagonalization of the Hamiltonian in each subspace
provides the necessary  information to calculate electromagnetic
transition rates. We will be interested here on the electric
quadrupole transitions, which apart from an unimportant scale
factor, are described by the quadrupole operator (\ref{Q}).
The action of this operator on the
basis states without broken pairs is

\begin{widetext}
\begin{equation}
\hat Q_{\mu} \vert \tilde n_{s} \tilde n_{d},0,0\rangle
=\frac{1}{\sqrt{C_{s,0}^{\tilde n_s}C_{d,0}^{\tilde n_d}}}
\left[\tilde n_{s}\left(
    K_{s}^{+}\right)
^{\tilde n_{s}-1}\left(  K_{d}^{+}\right)  ^{\tilde
  n_{d}}s^{\dagger} d^{\dagger}_\mu \left\vert 0\right\rangle
+ \tilde n_{d}\left(  K_{s}^{+}\right)  ^{\tilde n_{s}}\left(
K_{d}^{+}\right)  ^{\tilde n_{d}-1}s^{\dagger} d^{\dagger}_\mu\left\vert
0\right\rangle \right] ,
\end{equation}
\end{widetext}
where $\vert 0\rangle$ stands for the boson vacuum.

The matrix elements of interest if one wants to evaluate transition
rates from the ground state to the first excited state are%
\begin{equation}
\left\langle \left(  \tilde n_{s}-1\right)
\tilde n_{d},1,1\right\vert \hat Q_{\mu}\left\vert
\tilde n_{s}\tilde n_{d},0,0\right\rangle
=\sqrt{\frac{2\tilde n_{s}\left(  2\tilde n_{d}+5\right)}{5}},%
\end{equation}
\begin{equation}
\left\langle \tilde n_{s}\left(  \tilde n_{d}-1\right)
  ,1,1\right\vert \hat Q_{\mu}\left\vert
\tilde n_{s}\tilde n_{d},0,0\right\rangle
=\sqrt{\frac{2\tilde n_{d}\left(  2\tilde n_{s}+1\right)}{5}}.%
\end{equation}

If we write the eigenstates as%
\begin{equation}
\left\vert \Psi,\nu_{s}\nu_{d}\right\rangle =\sum_{\tilde n_{s},\tilde
  n_{d}}\varsigma_{\tilde n_{s},\tilde n_{d}%
}^{\nu_{s}\nu_{d}}\left\vert \tilde n_{s}\tilde
  n_{d},\nu_{s}\nu_{d}\right\rangle ,
\end{equation}
the matrix element of the $\hat Q$ operator between the ground state, $\vert
  \Psi,00\rangle$, and the first excited state, $\vert
\Phi,11\rangle $, is%

\begin{figure}[h]
  \centering
  \includegraphics[width=12cm]{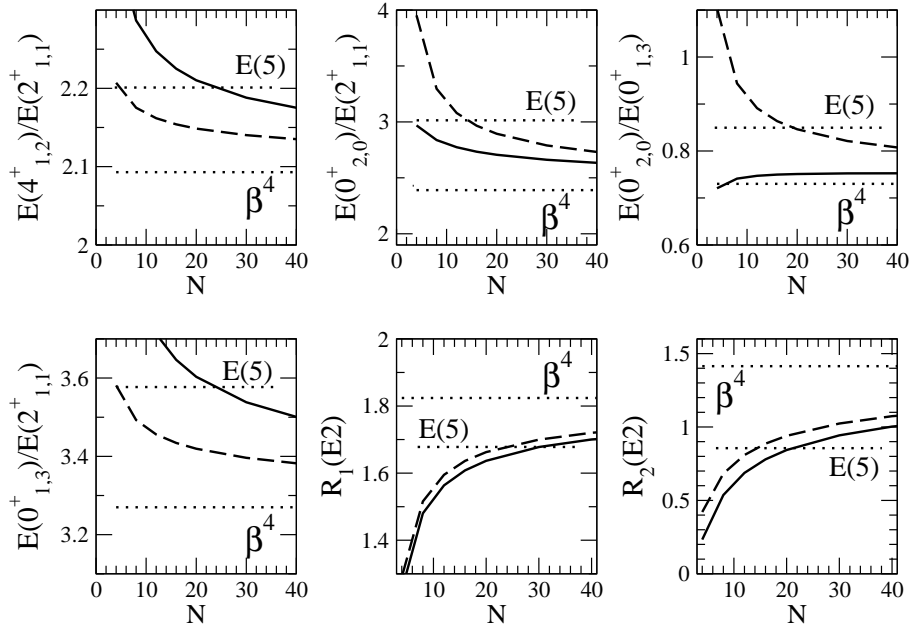}
\caption{Variation with the number of bosons (up to $N=40$) of
  selected energy and B(E2) ratios for IBM calculations
  performed at the critical points of Hamiltonians (\ref{HPP})
  (broken line) and (\ref{HQQ}) (full line). The corresponding E(5) and
  $\beta^4$ values are marked with horizontal dotted lines. }
\label{fig1}
\end{figure}

\begin{figure}[h]
  \centering
  \includegraphics[width=12cm]{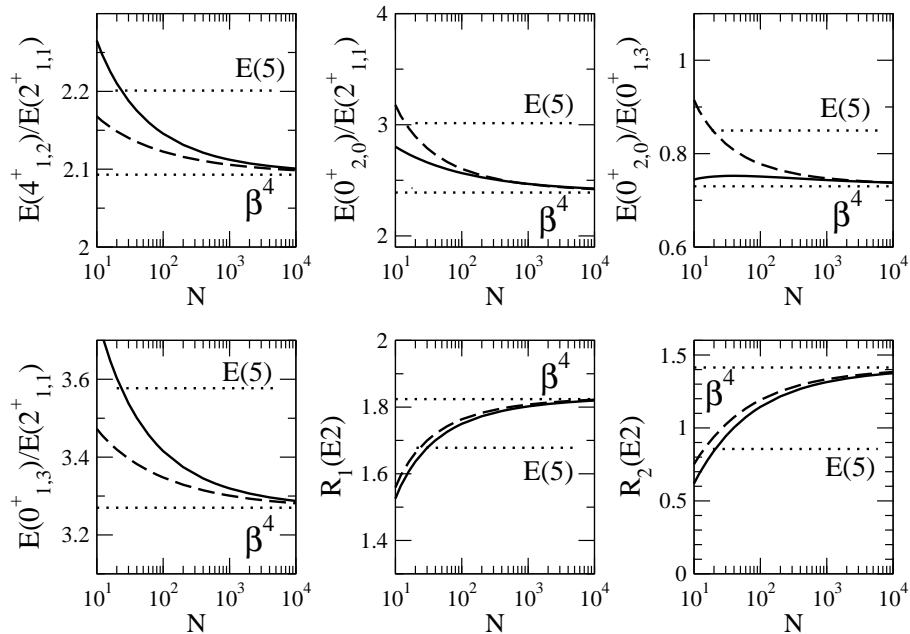}
\caption{Same as Fig. \ref{fig1} but here the number of bosons runs up
  to 10000 in both the energy and the B(E2) ratios. Please note
  the logarithmic scale in the number of boson axis.}
\label{fig2}
\end{figure}

\begin{equation}
\left\langle \Phi,11\right\vert \hat Q_{\mu}\left\vert \Psi,00\right\rangle
= \sum_{\tilde n_{s},\tilde n_{d}}
\left[\sqrt{\frac{2\tilde n_{s}\left(  2\tilde n_{d}+5\right)  }{5}%
}\varsigma_{\tilde n_{s},\tilde n_{d}}^{00}\varsigma_{\tilde n_{s}-1,\tilde n_{d}}^{11} \right.
+ \left. \sqrt{\frac{2\tilde n_{d}\left(
2\tilde n_{s}+1\right)  }{5}}\varsigma_{\tilde n_{s},\tilde
n_{d}}^{00}\varsigma_{\tilde n_{s},\tilde n_{d-1}}^{11} \right] .%
\end{equation}
The matrix elements of the electric quadrupole operator between
the first excited state ($\nu_s=1, \nu_d=1$) and the states with
$\nu_s=0, \nu_d=2$ can be calculated in a similar way.

In Fig. 2 we present some selected low energy eigenvalues and
$B(E2)$ ratios for boson numbers up to $N=40$ at the critical
points of both IBM Hamiltonians (\ref{HPP}) and (\ref{HQQ}). We
would like to emphasize here that the critical points for the two
Hamiltonians are different. The four energy ratios presented are
written explicitly in the figure and the two displayed B(E2)
ratios are: $R_1=\frac{B(E2;4^+_{1,2} \rightarrow
2^+_{1,1})}{B(E2;2^+_{1,1}
     \rightarrow
     0^+_{1,0})}$ and  $R_2=\frac{B(E2;0^+_{2,0} \rightarrow
    2^+_{1,1})}{B(E2;2^+_{1,1} \rightarrow 0^+_{1,0})}$.
Where we are using the notation $L^\pi_{\xi,\tau}$ to indicate the
states. The purpose of this figure is to correct a mistake we had
in Fig.~3 of Ref.~\cite{beta4} where the results for the
Hamiltonian (\ref{HQQ}) were calculated with a wrong value for
$x_c$. As can be seen in the figure, there are sizable differences
in the spectrum and transition rates between both Hamiltonians at
the critical points. Though from Fig. 2 a general tendency for
convergence to the solution of the Bohr equation with a $\beta^4$ potential
rather than to the E(5) symmetry is observed, the results, especially from the
B(E2)'s, are not yet conclusive. In Fig. 3 we show the new results
of this report. The same quantities as in Fig. 2 are plotted for
$N$ values up to 10000, including transition rates. In Ref.
\cite{beta4} energy eigenvalues were calculated up to $N=1000$ and
transition rates up to $N=40$. We can now clearly appreciate, both
from the energies and B(E2) transitions, that the IBM Hamiltonians
at the U(5) to O(6) critical point in the large $N$ limit converge
to the Bohr Hamiltonian with a $\beta^4$ potential.

\bigskip

In this report we make use of the property that the IBM Hamiltonian
along the transitional line from U(5) to O(6) is block diagonal
wit respect to the boson seniority quantum number and tridiagonal
within each subspace. This reduction allows to obtain exact
solutions up to very large number of bosons for energies and wave
functions. We have applied this formalism to confirm previous
studies about the correspondence between the IBM Hamiltonians at
the critical point in the U(5)--O(6) transition and the Bohr
Hamiltonian with a $\beta^4$ potential for the low energy
properties. This issue has also been studied recently from a
different point of view by Rowe et al. \cite{Rowe04}. The
formalism presented here for the IBM can be easily generalized to
other two-level boson models \cite{Dusuel}.

\section*{Acknowledgements}

This work was supported in part
by the Spanish DGICYT under projects number BFM2002-03315,
BFM2003-05316-C02-02,  BFM2003-05316, and FPA2003-05958.

%\bigskip

\end{document}